\documentclass[12pt]{article}
\usepackage{epsf}
\usepackage{graphicx}
\usepackage{amssymb}
\def\uu{\langle \bar u u \rangle}
\def\dd{\langle \bar d d \rangle}
\def\ss{\langle \bar s s \rangle}
\def\qq{\langle \bar q q \rangle}

\title{ {  Magnetic Dipole, Electric Quadrupole and Magnetic Octupole Moments of the $\Delta$ Baryons in  Light Cone QCD Sum Rules }}
\author{
  K. Azizi\thanks {e-mail:e146342@metu.edu.tr}
\\
  \small Physics
Department, Middle East Technical University, 06531, Ankara, Turkey }
 \date{}
\begin{document}
\setlength{\baselineskip}{26pt} \maketitle
\setlength{\baselineskip}{7mm}
\begin{abstract}
 Due to the very short life time of the $\Delta$ baryons,
 a direct measurement on the electromagnetic moments of these systems is almost impossible in the experiment and can only be done indirectly.
 Although only for the magnetic dipole moments of $\Delta^{++}$ and $\Delta^{+}$ systems there are some experimental data, the theoretical,
 phenomenological and lattice calculations could play crucial role. In  present work,   the
 magnetic dipole ($\mu_{\Delta}$) , electric quadrupole ($Q_{\Delta}$) and magnetic octupole ($O_{\Delta}$) moments of these baryons are
 computed within the  light cone QCD sum rules. The results are compared with the predictions of the other phenomenological approaches,
 lattice QCD and existing experimental data.
\end{abstract}
PACS: 11.55.Hx, 13.40.Em, 13.40.Gp, 14.20.-c
\thispagestyle{empty}
\newpage
\setcounter{page}{1}
\section{Introduction}
Study of the electromagnetic properties of the baryons can give
valuable information on their internal structure. Some of  the main
static electromagnetic parameters of the $\Delta$ baryons are their
magnetic dipole ($\mu_{\Delta}$), electric quadrupole ($Q_{\Delta}$)
and magnetic octupole ($O_{\Delta}$) moments. The
$\Delta^{-,+,++,0}$ baryons are the lowest and  very well-known
nucleon resonances. Because of their too short mean life time
($\sim10^{-23}$ s), there is almost no direct experimental
information about their form factors and electromagnetic moments. An
indirect measurement  for the magnetic dipole moment of
$\Delta^{++}$  was accurately done from the radiative pion-nucleon
scattering \cite{Nefkens} (see \cite{Yao} for experimental values of
the magnetic dipole moment of $\Delta^{++}$ obtained from various
experiments). The magnetic moment of the $\Delta^{+}$ resonance has
also been measured via $\gamma P\rightarrow \pi^0 \gamma'P$ reaction
in \cite{ Kotulla}.

 The magnetic dipole moments of these baryons have been studied in the framework of the various theoretical approaches.
 The radiative pion production on the nucleon ($\gamma N\rightarrow \pi N\gamma'$) with the aim of the determination of the
 magnetic dipole moment of the $\Delta^{+}(1232)$ has been studied in the frame work of Chiral effective- field theory in \cite{pascalutsa}.
 The magnetic dipole moment for $\Delta$ baryons is calculated in the framework of the  static quark model (SQM) \cite{Chiang},
 relativistic quark models (RQM) \cite{Schlumpf}, QCD sum rules
 (QCDSR) \cite{Lee1,AlievD}, Chiral quark-soliton models (ChQSM) \cite{Kim}, heavy baryon Chiral perturbation theory
(HBChPT) \cite{Butler,iki}, a phenomenological quark model (PQM)
which nonstatic effects of pion exchange and orbital excitation are
included \cite{franclin}, Lattice QCD
\cite{Leinweber,Cloet,Lee2,Alexandrou} and Chiral effective-field
theory \cite{bir}. The magnetic dipole, electric quadrupole and
magnetic octupole moments of these baryons is also calculated in
\cite{Pena} in the spectator quark formalism based on a simple
$\Delta$ wave function corresponding to a quark-diquark system in an
S-state. In \cite{Buchmann2}, the Quadrupole Moment of the $\Delta$
baryons are calculated in the frame work of the consistituent quark
model.   Recently, the octupole moments of the light decuplet
baryons are reported in \cite{Buchmann} within the non-covariant
quark model.

In the present work, we study the  magnetic dipole, electric
quadrupole and magnetic octupole  moments of the $\Delta$ baryons in
light cone QCD sum rules (LCSR) approach. Note that, by calculating
the electromagnetic form factors, the electromagnetic dipole moments
of the nucleons have been studied in \cite{kazem1} in the same frame
work. The paper contains 3 sections. In section 2, the light cone
QCD sum rules for the  magnetic dipole, electric quadrupole and
magnetic octupole  moments are calculated. Section 3 is devoted to
the numerical analysis of the  sum rules, a comparison of our
results with the  predictions of the other approaches as well as the
existing experimental data   and also discussion.
\section{Light cone QCD sum rules for  the magnetic dipole, electric quadrupole and magnetic octupole moments  of the $\Delta$ baryons  }
To calculate   the magnetic dipole,  electric quadrupole and
magnetic octupole moments  of the $\Delta$ baryons, we start
considering the basic object in LCSR method  (the correlation
function),  where hadrons are represented by the interpolating quark
currents.
\begin{equation}\label{T}
T_{\mu\nu}=i\int d^{4}xe^{ipx}\langle0\mid T\{\eta_{\mu}(x)\bar{\eta}_{\nu}(0) \}\mid0\rangle_{\gamma},
\end{equation}
where $\eta_{\mu}$ is the interpolating current of the $\Delta$ baryons
and $\gamma$ stands for  the electromagnetic field. In QCD sum rules approach, this
correlation function is calculated in two different languages: in the
quark-gluon language (QCD or theoretical side), it describes a hadron as quarks and gluons interacting
in QCD vacuum. In the physical side, it is saturated by complete sets
of hadrons with the same  quantum numbers as their interpolating currents. The physical quantities, i.e.,  the electromagnetic form factors and multipole moments are
calculated  equating these  two different representations of the
correlation function.

The physical or phenomenological side of the correlation function
can be obtained  inserting the complete sets of the hadronic states
between the interpolating currents in  Eq. (\ref{T}) with the same
quantum numbers as the $\Delta$ baryons, i.e.,
\begin{eqnarray}\label{T2}
T_{\mu\nu}&=&\frac{\langle0\mid \eta_{\mu}\mid
\Delta(p_{2})\rangle}{p_{2}^{2}-m_{\Delta}^{2}}\langle \Delta(p_{2})\mid
\Delta(p_{1})\rangle_\gamma\frac{\langle \Delta(p_{1})\mid
\bar{\eta}_{\nu}\mid 0\rangle}{p_{1}^{2}-m_{\Delta}^{2}},
\end{eqnarray}
where $p_{1}=p+q$,  $p_{2}=p$ and q is the momentum of the photon. The matrix element of the interpolating current between the vacuum and the baryon state is defined as
\begin{equation}\label{lambdabey}
\langle0\mid \eta_{\mu}(0)\mid \Delta(p,s)\rangle=\lambda_{\Delta}u_{\mu}(p,s),
\end{equation}
where $\lambda_{\Delta}$ is the  residue and $u_{\mu}(p,s)$ is the
Rarita-Schwinger spinor. The matrix element $\langle
\Delta(p_{2})\mid \Delta(p_{1})\rangle_\gamma$ entering Eq.
(\ref{T2}) can be parameterized in terms of some form factors as
\cite{Pena, Pascalutsa07,Nozawa90}:
\begin{eqnarray}\label{matelpar}
\langle \Delta(p_{2})\mid \Delta(p_{1})\rangle_\gamma &=&-e\bar
u_{\mu}(p_{2})\left\{\vphantom{\int_0^{x_2}}F_{1}g^{\mu\nu}\not\!\varepsilon-
\frac{1}{2m_{\Delta}}\left
[F_{2}g^{\mu\nu}+F_{4}\frac{q^{\mu}q^{\nu}}{(2m_{\Delta})^2}\right]\not\!\varepsilon\not\!q
\right.\nonumber\\&+&\left.
F_{3}\frac{1}{(2m_{\Delta})^2}q^{\mu}q^{\nu}\not\!\varepsilon\vphantom{\int_0^{x_2}}\right\} u_{\nu}(p_{1}),\nonumber\\
\end{eqnarray}
where $\varepsilon$ is the polarization vector of the photon  and  $F_{i}$ are the form factors as functions of $q^2=(p_{1}-p_{2})^2$. In obtaining  the  expression for  the correlation function,  summation over spins of the $\Delta$ particles is performed using
\begin{equation}\label{raritabela}
\sum_{s}u_{\mu}(p,s)\bar u_{\nu}(p,s)=\frac{(\not\!p+m_{\Delta})}{2m_{\Delta}}\{-g_{\mu\nu}
+\frac{1}{3}\gamma_{\mu}\gamma_{\nu}-\frac{2p_{\mu}p_{\nu}}
{3m^{2}_{\Delta}}-\frac{p_{\mu}\gamma_{\nu}-p_{\nu}\gamma_{\mu}}{3m_{\Delta}}\}.
\end{equation}
In deriving the  expression for   the phenomenological side of the correlation function appear two problems (see also \cite{kazem2}): 1) all Lorentz structures are not independent, 2) not only spin 3/2, but spin 1/2 states  also contribute to the correlator, which should be eliminated. Indeed, the matrix element of the current $\eta_{\mu}$ between vacuum and spin 1/2 states is nonzero and is determined as
\begin{equation}\label{spin12}
\langle0\mid \eta_{\mu}(0)\mid B(p,s=1/2)\rangle=(A  p_{\mu}+B\gamma_{\mu})u(p,s=1/2).
\end{equation}
Imposing the condition $\gamma_\mu \eta^\mu = 0$, one can
immediately obtain that $B=-\frac{A}{4}m$.

To remove the spin 1/2 contribution and obtain only independent structures in the
correlation function, we order Dirac matrices  in a specific form.
For this aim, we choose the  ordering for Dirac
matrices as $\gamma_{\mu}\not\!p\not\!\varepsilon\not\!q\gamma_{\nu}$. With this ordering for the correlator, we obtain
\begin{eqnarray}\label{final phenpart}
T_{\mu\nu}&=&-\lambda_{_{\Delta}}^{2}\frac{1}{(p_{1}^{2}-m_{_{\Delta}}^{2})(p_{2}^{2}-m_{_{\Delta}}^{2})}
\left[\vphantom{\int_0^{x_2}}2m_{\Delta}(\varepsilon.p)g_{\mu\nu}F_{1}\right.
\nonumber\\&+&\frac{1}{m_{\Delta}}(\varepsilon.p)g_{\mu\nu}\not\!p\not\!qF_{2}+
\frac{1}{2m_{\Delta}^2}(\varepsilon.p)q_{\mu}q_{\nu}\not\!pF_{3}
\nonumber\\&+&\frac{1}{4m_{\Delta}^2}(\varepsilon.p)q_{\mu}q_{\nu}\not\!qF_{4}+\mbox{other independent structures  }\nonumber\\&+&\mbox{ structures with }\gamma_{\mu}\mbox{at the beginning and} \gamma_{\nu}\mbox{ at the end } \nonumber\\
&&\mbox{  or
which are proportional to }p_{2\mu}\mbox{or }p_{1\nu}\left.\vphantom{\int_0^{x_2}}\right].
\end{eqnarray}
The magnetic dipole ($G_{m}(q^2)$), electric quadrupole
$(G_{Q}(q^2)$) and magnetic octupole $(G_{O}(q^2)$) form factors are
defined in terms of the form factors $F_{i}(q^2)$ in the following
way
 \cite{Pena, Pascalutsa07,Nozawa90,Weber78}:
\begin{eqnarray}
G_{m}(q^2) &=& \left[ F_1(q^2) + F_2(q^2)\right] ( 1+ \frac{4}{5}
\tau ) -\frac{2}{5} \left[ F_3(q^2)  +
F_4(q^2)\right] \tau \left( 1 + \tau \right) \nonumber\\
G_{Q}(q^2) &=& \left[ F_1(q^2) -\tau F_2(q^2) \right]  -
\frac{1}{2}\left[ F_3(q^2) -\tau F_4(q^2)
\right] \left( 1+ \tau \right)  \nonumber \\
 G_{O}(q^2) &=&
\left[ F_1(q^2) + F_2(q^2)\right] -\frac{1}{2} \left[ F_3(q^2)  +
F_4(q^2)\right] \left( 1 + \tau \right),\end{eqnarray}
  where $\tau
= -\frac{q^2}{4m_\Delta^2}$. At $q^2=0$, the multipole form factors
are obtained in terms of the functions $F_i(0)$ as:
\begin{eqnarray}\label{mqo1}
G_{m}(0)&=&F_{1}(0)+F_{2}(0)\nonumber\\
G_{Q}(0)&=&F_{1}(0)-\frac{1}{2}F_{3}(0)\nonumber\\
G_{O}(0)&=&F_{1}(0)+F_{2}(0)-\frac{1}{2}[F_{3}(0)+F_{4}(0)].
\end{eqnarray}
The  static  magnetic dipole ($\mu_{\Delta}$), electric quadrupole
($Q_{\Delta}$)  and magnetic octupole ($O_{\Delta}$) moments  in
their natural  magneton are defined in the following way:
 \begin{eqnarray}\label{mqo2}
\mu_{\Delta}&=&\frac{e}{2m_{\Delta}}G_{m}(0)\nonumber\\
Q_{\Delta}&=&\frac{e}{m_{\Delta}^2}G_{Q}(0)\nonumber\\
O_{\Delta}&=&\frac{e}{2m_{\Delta}^3}G_{O}(0).
\end{eqnarray}

The theoretical part of the correlation function can be calculated
in light cone QCD sum rules via the operator product expansion (OPE)
in deep Euclidean region where $p^2\ll0$ and $(p+q)^2\ll0$ in terms
of the photon distribution amplitudes (DA's). To calculate the
correlation function from theoretical or QCD side, the explicit
expressions of the interpolating currents of the $\Delta$ baryons
are needed. The interpolating current for $\Delta^+$  is chosen as
\begin{eqnarray}\label{currentguy}
\eta_{\mu}=\frac{1}{\sqrt{3}}\varepsilon^{abc}\left[\vphantom{\int_0^{x_2}}2(u^{aT}C\gamma_{\mu}d^{b})u^{c}+(u^{aT}C\gamma_{\mu}u^{b})d^{c}\right],
\end{eqnarray}
where C is the charge conjugation operator and  a, b and c are color
indices. Here we should mention that in the present work, first we calculate the correlation function for $\Delta^+$ then with the help of the relations which we will present next, the correlators of $\Delta^-$, $\Delta^{++}$ and $\Delta^0$ will be obtained using the  correlation function of the $\Delta^+$.
After contracting out the quark pairs in  Eq. (\ref{T}) by the help of the
Wick's theorem, we obtain the following expression for the correlation
function in terms of the  quark propagators
\begin{eqnarray}\label{tree expresion.m}
\Pi_{\mu\nu}&=&\frac{i}{3}\epsilon_{abc}\epsilon_{a'b'c'}\int
d^{4}xe^{ipx}\langle0[\gamma(q)]\mid\{2S_{d}^{ca'}
\gamma_{\nu}S'^{ab'}_{u}\gamma_{\mu}S_{u}^{bc'}\nonumber\\&-&2S_{d}^{ca'}
\gamma_{\nu}S'^{bb'}_{u}\gamma_{\mu}S_{u}^{ac'}+ 2S_{u}^{ca'}
\gamma_{\nu}S'^{ab'}_{u}\gamma_{\mu}S_{d}^{bc'}-2S_{u}^{cb'}
\gamma_{\nu}S'^{aa'}_{u}\gamma_{\mu}S_{d}^{bc'}\nonumber\\&+&4 S_{u}^{cb'}
\gamma_{\nu}S'^{ba'}_{d}\gamma_{\mu}S_{u}^{ac'}
+Tr(\gamma_{\mu}S_{u}^{aa'}\gamma_{\nu}S'^{bb'}_{u})S^{cc'}_{d}
\nonumber\\&-&Tr(\gamma_{\mu}S_{u}^{ab'}\gamma_{\nu}S'^{ba'}_{u})S^{cc'}_{d}-4
Tr(\gamma_{\mu}S_{u}^{ab'}\gamma_{\nu}S'^{ba'}_{d})S^{cc'}_{u}\}\mid 0\rangle,
\end{eqnarray}
where $S'=CS^TC$ and $S_{u,d}$ are the full light quark
propagators, which their explicit expressions can be found in \cite{Balitsky, Braun2} (see also \cite{kazem2, kabolishokolati}). To calculate the above correlation function, we follow the same procedure as stated in \cite{kazem2, kabolishokolati} and use
 the photon distribution amplitudes (DA's) calculated in  \cite{Ball}. For convenience, we  present those DA's in the appendix--A.

Using the expressions of the full light    propagator and the photon
DA's and separating the coefficient of the structures
$(\varepsilon.p)g_{\mu\nu}$,
$(\varepsilon.p)g_{\mu\nu}\not\!p\not\!q$, $
(\varepsilon.p)q_{\mu}q_{\nu}\not\!p$ and  $
(\varepsilon.p)q_{\mu}q_{\nu}\not\!q$ for the  $F_{1}$, $F_{2}$,
$F_{3}$ and $F_{4}$, respectively, the expressions of the
correlation function from the QCD side are  obtained. Separating the
coefficient of the same structures from phenomenological part and
equating these representations of the correlator,  sum rules for the
 $F_i$ functions are obtained. In order to
suppress the contribution of the higher states and continuum, Borel
transformation with respect to the variables $p_{2}^2=p^2$ and
$p_{1}^2=(p+q)^2$
 is applied. The explicit expressions for $F_{i}$ are given in the appendix--B.

At the end of this section, we would like to present some  relations
between the correlation functions. Our calculations show that the
coefficient of any structure in the correlation function of the
$\Delta^+$ can be written in the form \begin{eqnarray}\label{relal1}
\Pi^{\Delta^{+}} &=& - \frac{1}{6} (2 e_u + e_d) {\cal H}(u,d),\nonumber \\
\end{eqnarray}
where the function ${\cal H}(u,d)$ depends on the masses and condensates of the u and d quarks and it is  independent of the charge of the quarks. Our calculations indicate that the $\Pi^{\Delta^{-,0,++}}$  can be obtained from the $\Pi^{\Delta^{+}}$ by the following replacements:
\begin{eqnarray}\label{relal1}
\Pi^{\Delta^{++}} &=&\Pi^{\Delta^{+}}(d\rightarrow u)= - \frac{1}{2} e_u {\cal H}(u,u), \nonumber \\
\Pi^{\Delta^{0}} &=& \Pi^{\Delta^{+}}(d\leftrightarrow u)= - \frac{1}{6} (2 e_d + e_u) {\cal H}(d,u), \nonumber \\
\Pi^{\Delta^{-}} &=& \Pi^{\Delta^{+}}(u\rightarrow d)=- \frac{1}{2} e_d {\cal H}(d,d),
\end{eqnarray}
We  consider the  massless quarks, $m_u = m_d = 0$, and exact SU(2)
flavor symmetry implying $\langle \bar u u \rangle = \langle \bar d
d \rangle$.  Under exact $SU(2)$ flavor symmetry, ${\cal
H}(u,d)={\cal H}(d,u)={\cal H}(u,u)={\cal H}(d,d) ={\cal H}$ and the
following relations are obtained (see also \cite{AlievD}):
\begin{eqnarray}\label{relal1}
\Pi^{\Delta^{++}} &=& - \frac{1}{2} e_u {\cal H}, \nonumber \\
\Pi^{\Delta^{+}} &=& - \frac{1}{6} (2 e_u + e_d) {\cal H} ,\nonumber \\
\Pi^{\Delta^{0}} &=& - \frac{1}{6} (2 e_d + e_u) {\cal H}, \nonumber \\
\Pi^{\Delta^{-}} &=& - \frac{1}{2} e_d {\cal H},
\end{eqnarray}
 From above equation, by substituting the charge of the u and d quarks, the following exact
relations between theoretical parts of the correlator of $\Delta$ baryons are derived:
\begin{eqnarray} \label{relal2}
\Pi^{\Delta^+} &=& - \Pi^{\Delta^-} = \frac{1}{2} \Pi^{\Delta^{++}} \nonumber
\\
\Pi^{\Delta^0} &=& 0 \label{relal}
\end{eqnarray}

%

\section{Numerical analysis}
Present  section encompasses the numerical analysis for the,
magnetic dipole, electric quadrupole and magnetic octupole  moments
of the $\Delta$  baryons. The values for input parameters used in
the analysis of the sum rules for the $F_{1}$, $F_{2}$, $F_{3}$ and
$F_{4}$ are : $\uu(1~GeV) = \dd(1~GeV)= -(0.243)^3~GeV^3$,
$\ss(1~GeV) = 0.8 \uu(1~GeV)$, $m_0^2(1~GeV) = (0.8\pm0.2)~GeV^2$
\cite{Belyaev} and $f_{3 \gamma} = - 0.0039~GeV^2$ \cite{Ball}. The
value of the magnetic susceptibility was obtained  in various papers
as $\chi(1~GeV)=-3.15\pm0.3~GeV^{-2}$  \cite{Ball},
$\chi(1~GeV)=-(2.85\pm0.5)~GeV^{-2}$ \cite{Rohrwild} and
$\chi(1~GeV)=-4.4~GeV^{-2}$\cite{Kogan}. The residue
$\lambda_{\Delta} $ determined from mass sum rules and is taken to
be $\lambda_{\Delta} = 0.038~ GeV ^{3}$ \cite{Lee1,Hwang,kazem3}.
From sum rules for the  $F_{1}$, $F_{2}$, $F_{3}$ and $F_{4}$, it
follows that the photon DA's are also needed \cite{Ball}. Their
explicit expressions are also given in the appendix--A.

The sum rules for the magnetic dipole, electric quadrupole and
magnetic octupole  moments also contain two auxiliary parameters:
Borel mass parameter $M^2$ and continuum threshold $s_{0}$. The
physical quantities, i.e.,  magnetic dipole, electric quadrupole and
magnetic octupole moments,  should be independent of these
parameters. The working region for  $M^2$  is determined requiring
that the contributions of the higher states and continuum are
effectively suppressed. This condition is satisfied in the region
$1~GeV^2\leq M^{2}\leq1.5~GeV^2 $.
\begin{figure}[h!]
\begin{center}
\includegraphics[width=9cm]{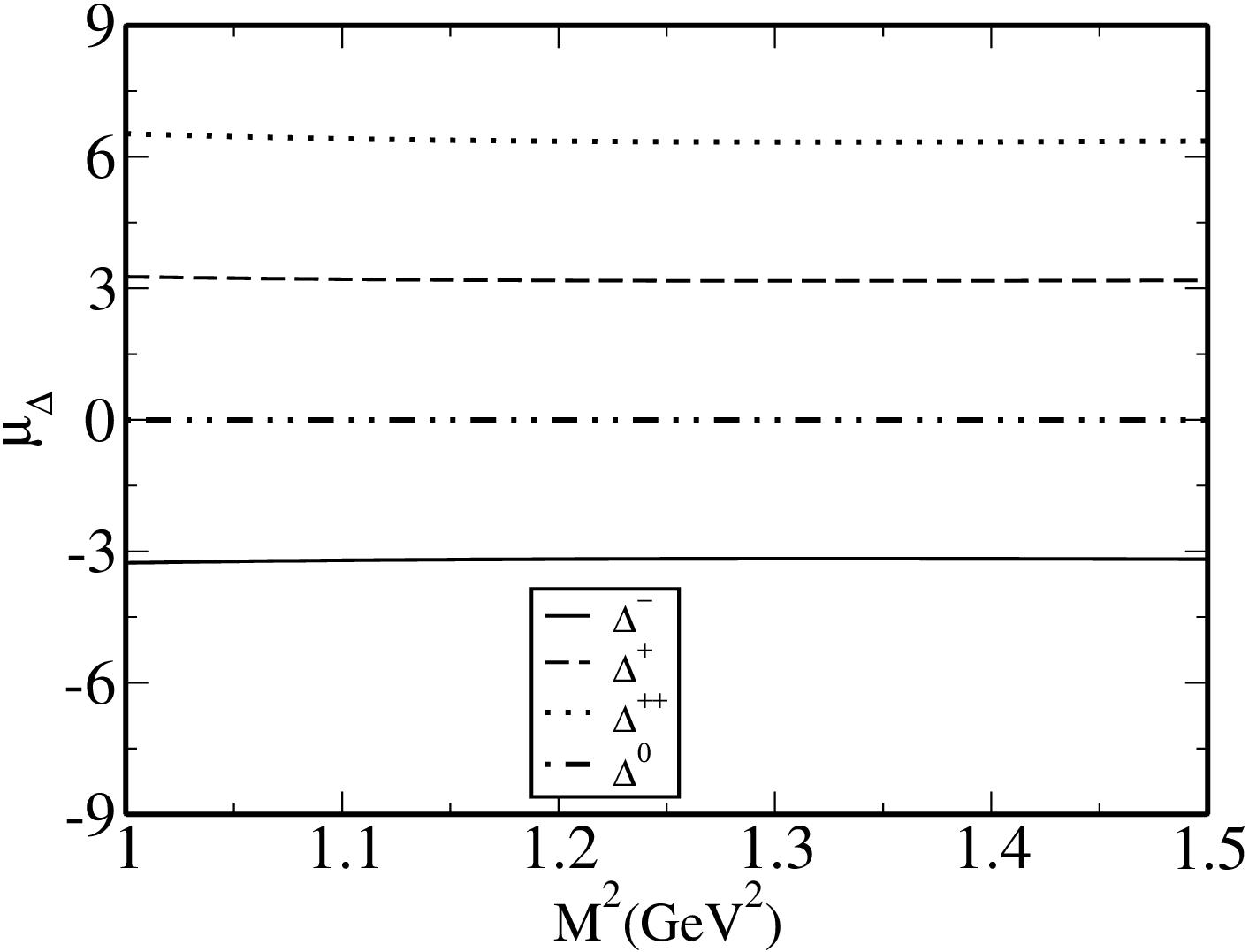}
\end{center}
\caption{The dependence of the magnetic dipole moment $\mu_{\Delta}$
in its natural  magneton   on the Borel parameter $M^{2}$ at
 fixed value of the continuum
threshold $s_{0}=4~GeV^2$.} \label{fig2}
\end{figure}
\begin{figure}[h!]
\begin{center}
\includegraphics[width=9cm]{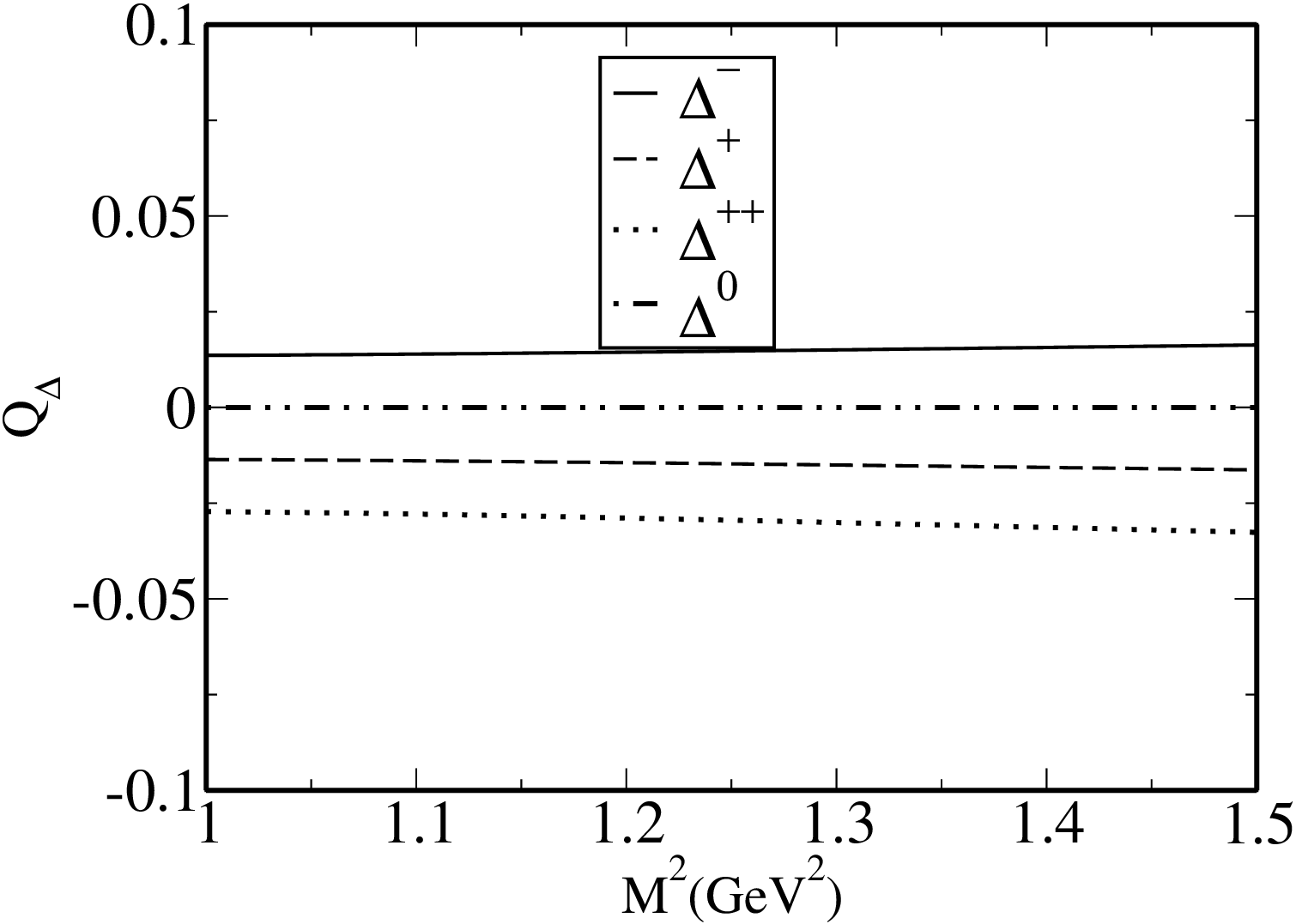}
\end{center}
\caption{The dependence of the electric quadrupole  $Q_{\Delta}$ in
$fm^2$ on the Borel parameter $M^{2}$
 at fixed value of the continuum
threshold $s_{0}=4~GeV^2$.  } \label{fig4}
\end{figure}
\begin{figure}[h!]
\begin{center}
\includegraphics[width=9cm]{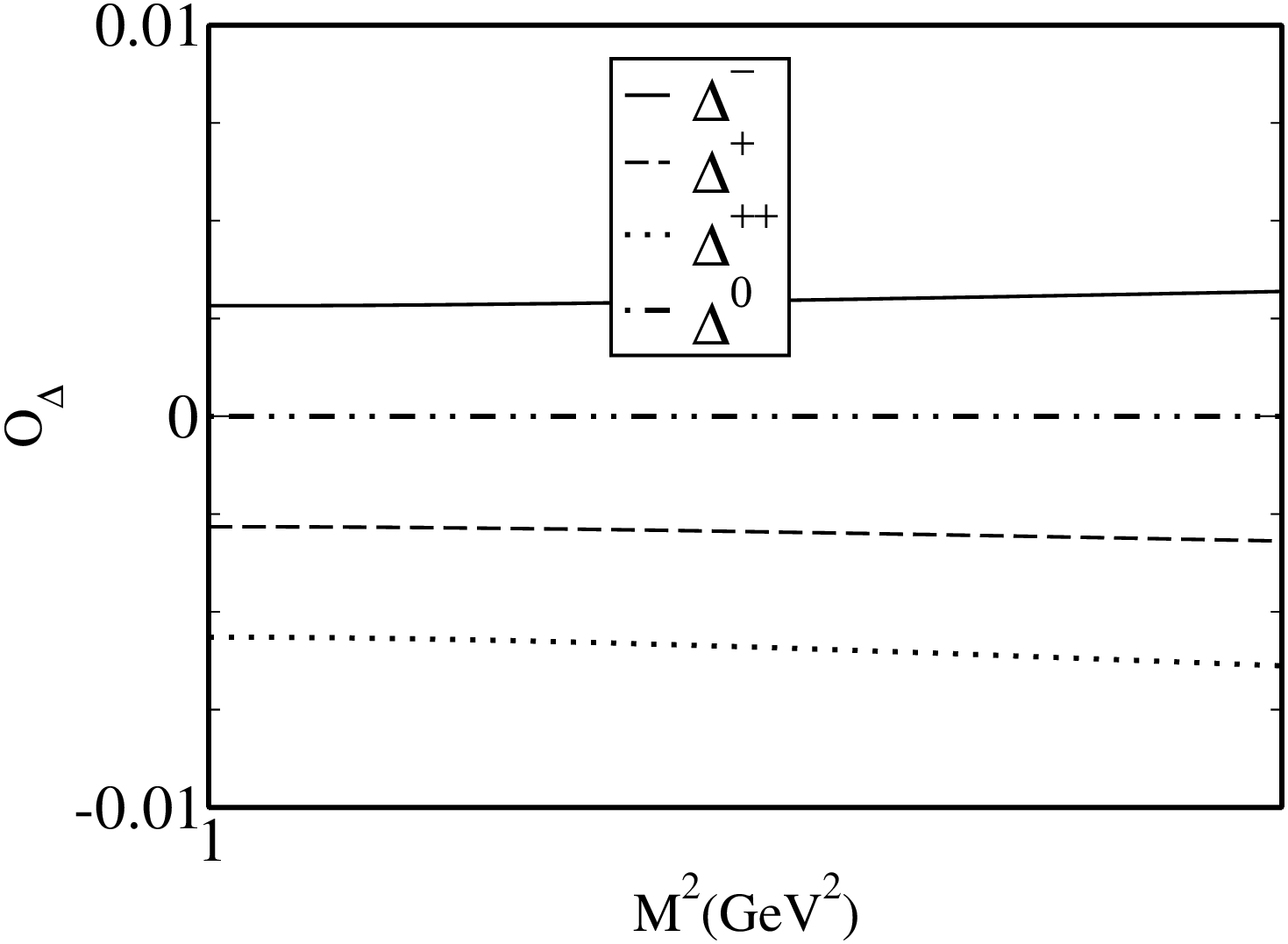}
\end{center}
\caption{The same as Fig. 4, but for the magnetic octupole
$O_{\Delta}$ in $fm^3$.  } \label{fig5}
\end{figure}
\begin{table}[h!]
\centering
\begin{tabular}{|c||c|c|c|c|}\hline
 &$\Delta^{-}$ & $\Delta^{+}$& $\Delta^{++}$ &$\Delta^{0}$ \\\cline{1-5}
\hline\hline present work&$-3.17\pm0.85$&$3.17\pm0.85$
&$6.34\pm1.70$&$0$\\\cline{1-5} Exp.\cite{
Kotulla}&$-$&$3.54^{+1.3}_{-1.7}\pm1.96\pm3.93$ &-&$-$\\\cline{1-5}
Exp.\cite{Yao}&-&- &$7.34\pm2.49$&-\\\cline{1-5}
SQM\cite{Chiang}&-3.65 &3.65 &7.31&0\\\cline{1-5}
RQM\cite{Schlumpf}&-3.12&3.12&6.24&0\\\cline{1-5}
QCDSR\cite{Lee1}&$-2.88\pm0.52$&$2.88\pm0.52$&$5.76\pm1.05$&0\\\cline{1-5}
QCDSR\cite{AlievD}&$-2.71\pm0.85$&$2.71\pm0.85$&$5.41\pm1.70$&0\\\cline{1-5}
ChQSM\cite{Kim}&-3.69&3.48&7.06&-0.10\\\cline{1-5}
HBChPT\cite{Butler}&$-2.95\pm0.33$&$2.75\pm0.26$&$6.24\pm0.52$&$-0.22\pm0.05$\\\cline{1-5}
PQM\cite{franclin}&-&3.72&8.10 &-\\\cline{1-5}
Lattice\cite{Leinweber}&$-3.22\pm0.41$&$3.22\pm0.41$&$6.43\pm0.80$&0\\\cline{1-5}
Lattice\cite{Cloet}&$-3.22\pm0.35$&$3.26\pm0.35$&$6.54\pm0.73$&0.079\\\cline{1-5}
Lattice\cite{Lee2}&$-3.90\pm0.25$&$1.27\pm0.10$&$6.86\pm0.24$&$-0.046\pm0.003$\\\cline{1-5}
Lattice\cite{ Alexandrou}&-&$3.04\pm0.21$&-&-\\\cline{1-5}
Spectator\cite{ Pena}&-3.54&3.29&6.71&-0.12\\\cline{1-5}
 \end{tabular}
 \vspace{0.8cm}
\caption{Comparison of the magnetic dipole moment $\mu_{\Delta}$ in
units of its natural magneton  for different approaches like static
quark model (SQM) \cite{Chiang},  relativistic quark models (RQM)
\cite{Schlumpf}, QCD sum rules
 (QCDSR) \cite{Lee1,AlievD}, Chiral quark-soliton models (ChQSM) \cite{Kim}, heavy baryon Chiral perturbation theory
(HBChPT) \cite{Butler}, a phenomenological quark model (PQM) which
nonstatic effects of pion exchange and orbital excitation are
included \cite{franclin}, Lattice QCD
\cite{Leinweber,Cloet,Lee2,Alexandrou} and experiment
\cite{Yao,Kotulla}. The presented experimental value for
$\Delta^{++}$ is the average of sum data from \cite{Yao}.
}\label{tab:1}
\end{table}
\begin{table}[h!]
\centering
\begin{tabular}{|c||c|c|c|c|}\hline
 &$\Delta^{-}$ & $\Delta^{+}$& $\Delta^{++}$ &$\Delta^{0}$ \\\cline{1-5}
\hline\hline $Q_{\Delta}$(present work)&$0.014\pm0.004$
&$-0.014\pm0.004$ &$-0.028\pm0.008$&0\\\cline{1-5}
$Q_{\Delta}$\cite{Pena} &0 &0&0&0\\\cline{1-5}
$Q_{\Delta}^{imp}$\cite{Buchmann2} &0.032
&-0.032&-0.064&0\\\cline{1-5} $Q_{\Delta}^{exc}$\cite{Buchmann2}
&0.119 &-0.119&-0.238&0\\\cline{1-5}
\hline\hline$O_{\Delta}$(present
work)&$0.003\pm0.001$&$-0.003\pm0.001$&$-0.006\pm0.002$&0\\\cline{1-5}
$O_{\Delta}$\cite{Pena} &0 &0&0&0\\\cline{1-5}
$O_{\Delta}$\cite{Buchmann} &0.012 &-0.012&-0.024&0\\\cline{1-5}
 \end{tabular}
 \vspace{0.8cm}
\caption{Results for the  electric quadrupole $Q_{\Delta}$ in $fm^2$
and magnetic octupole $O_{\Delta}$ moments  in $fm^3$ in different
approaches: LCSR(present work), spectator quark model(\cite{Pena}),
constituent quark model  with configuration mixing but no exchange
currents (impulse approximation), constituent quark model  with
exchange currents but no configuration mixing \cite{Buchmann2} and
non-covariant quark model \cite{Buchmann}. }\label{tab:1}
\end{table}

 The dependency of the magnetic dipole moment $\mu_{\Delta}$,
   electric quadrupole $Q_{\Delta}$ and magnetic octupole $O_{\Delta}$ moments on Borel parameter $M^2$ are presented in Figs. 1-3
   at fixed value of the continuum threshold $s_{0}=4~GeV^2$. The
magnetic dipole moment is presented in its natural magneton ($
e\hslash /2m_{\Delta} c$) while the    electric quadrupole
($Q_{\Delta}$) and magnetic octupole ($O_{\Delta}$) moments are
shown in $fm^2$ and $fm^3$, respectively.  The conversion
coefficient from the natural magneton unit to the nucleon magneton
is $\frac{m_{N}}{m_{\Delta}}$. Note that, our results are
practically the same in the interval $s_{0}=(3.8-4.2)~GeV^2$ for
continuum threshold. These figures
  present a good stability with respect to the Borel mass parameter.  We should also
mention  that our results practically don't change considering three
values of the $\chi$ as presented at the beginning of this
section.

Our final  results on the magnetic dipole moment $\mu_{\Delta}$ for
$\Delta$ baryons are presented in Table 1. The quoted errors for the
values are due to the uncertainties in the determination of the
input parameters, the variation of $M^{2}$ as well as the systematic
errors in QCD sum rules approach. For comparison, the predictions of
other theoretical approaches, lattice QCD as well as the experiment
are also presented.
 From this Table, we see a good consistency among the various approaches especially
 when we consider the errors except the lattice QCD prediction \cite{Lee2} for magnetic moment of $\Delta^+$.

 We also depict the results of  the  electric quadrupole $Q_{\Delta}$ and magnetic octupole $O_{\Delta}$ moments in Table 2.
  In comparison, the results of the other approaches  are  also presented. From this Table, we see that the values for  the electric quadrupole  and magnetic octupole
 moments are very small in comparison with the magnetic dipole
 moment. Our results on  electric quadrupole moments  are consistent with the predictions of the constituent quark model
   with configuration mixing but no exchange currents (impulse approximation)\cite{Buchmann2} in order of magnitude,
   but about
    one order of magnitude smaller than the predictions of constituent quark model  with  exchange currents but no configuration mixing \cite{Buchmann2}.
     The \cite{ Pena}  predicts no electric quadrupole and magnetic octupole moments for
  $\Delta$ baryons. Our results on the magnetic octupole moments for these baryons are about four times smaller than the predictions of the
   non-covariant quark model \cite{Buchmann}.  The negative sign in the value of the quadrupole
    and octupole  moments of $\Delta^{+}$ shows  that the quadrupole and octupole distributions are oblate and have the
     same geometric shape as the charge distribution.

In conclusion, due to the very short life time, a direct measurement on the electromagnetic moments of $\Delta$
 systems is almost not possible in the experiment and can only be done indirectly in a three-step process, where
  they are created, emit a low-energy photon and then decay. Although only for $\Delta^{+}$ and $\Delta^{++}$ systems there are some
  data, the theoretical, phenomenological and lattice calculations could play very important role. In  present
   work,  we
   computed  the magnetic dipole, electric quadrupole  and magnetic octupole  moments of these baryons in the framework of the light cone QCD sum rules
    and compared their results
   with the predictions of the other phenomenological models, lattice QCD as well as the existing experimental data. The results depict that the
    electric quadrupole  and magnetic octupole  moments are very small in comparison with the  magnetic dipole
    moment of these baryons.
\section{Acknowledgment}
The author would like to thank T. M. Aliev and A. Ozpineci for their
useful discussions and also TUBITAK, Turkish Scientific and Research
Council, for their partial financial support.

\clearpage
\section*{Appendix A}
 The matrix elements used in the calculations are given in terms of the photon distribution amplitudes (DA's) as follows \cite{Ball}:
 \begin{eqnarray}
&&\langle \gamma(q) \vert  \bar q(x) \sigma_{\mu \nu} q(0) \vert  0
\rangle  = -i e_q \bar q q (\varepsilon_\mu q_\nu - \varepsilon_\nu
q_\mu) \int_0^1 du e^{i \bar u qx} \left(\chi \varphi_\gamma(u) +
\frac{x^2}{16} \mathbb{A}  (u) \right) \nonumber \\ &&
-\frac{i}{2(qx)}  e_q \qq \left[x_\nu \left(\varepsilon_\mu - q_\mu
\frac{\varepsilon x}{qx}\right) - x_\mu \left(\varepsilon_\nu -
q_\nu \frac{\varepsilon x}{q x}\right) \right] \int_0^1 du e^{i \bar
u q x} h_\gamma(u)
\nonumber \\
&&\langle \gamma(q) \vert  \bar q(x) \gamma_\mu q(0) \vert 0 \rangle
= e_q f_{3 \gamma} \left(\varepsilon_\mu - q_\mu \frac{\varepsilon
x}{q x} \right) \int_0^1 du e^{i \bar u q x} \psi^v(u)
\nonumber \\
&&\langle \gamma(q) \vert \bar q(x) \gamma_\mu \gamma_5 q(0) \vert 0
\rangle  = - \frac{1}{4} e_q f_{3 \gamma} \epsilon_{\mu \nu \alpha
\beta } \varepsilon^\nu q^\alpha x^\beta \int_0^1 du e^{i \bar u q
x} \psi^a(u)
\nonumber \\
&&\langle \gamma(q) | \bar q(x) g_s G_{\mu \nu} (v x) q(0) \vert 0
\rangle = -i e_q \qq \left(\varepsilon_\mu q_\nu - \varepsilon_\nu
q_\mu \right) \int {\cal D}\alpha_i e^{i (\alpha_{\bar q} + v
\alpha_g) q x} {\cal S}(\alpha_i)
\nonumber \\
&&\langle \gamma(q) | \bar q(x) g_s \tilde G_{\mu \nu} i \gamma_5 (v
x) q(0) \vert 0 \rangle = -i e_q \qq \left(\varepsilon_\mu q_\nu -
\varepsilon_\nu q_\mu \right) \int {\cal D}\alpha_i e^{i
(\alpha_{\bar q} + v \alpha_g) q x} \tilde {\cal S}(\alpha_i)
\nonumber \\
&&\langle \gamma(q) \vert \bar q(x) g_s \tilde G_{\mu \nu}(v x)
\gamma_\alpha \gamma_5 q(0) \vert 0 \rangle = e_q f_{3 \gamma}
q_\alpha (\varepsilon_\mu q_\nu - \varepsilon_\nu q_\mu) \int {\cal
D}\alpha_i e^{i (\alpha_{\bar q} + v \alpha_g) q x} {\cal
A}(\alpha_i)
\nonumber \\
&&\langle \gamma(q) \vert \bar q(x) g_s G_{\mu \nu}(v x) i
\gamma_\alpha q(0) \vert 0 \rangle = e_q f_{3 \gamma} q_\alpha
(\varepsilon_\mu q_\nu - \varepsilon_\nu q_\mu) \int {\cal
D}\alpha_i e^{i (\alpha_{\bar q} + v \alpha_g) q x} {\cal
V}(\alpha_i) \nonumber \\ && \langle \gamma(q) \vert \bar q(x)
\sigma_{\alpha \beta} g_s G_{\mu \nu}(v x) q(0) \vert 0 \rangle  =
e_q \qq \left\{
        \left[\left(\varepsilon_\mu - q_\mu \frac{\varepsilon x}{q x}\right)\left(g_{\alpha \nu} -
        \frac{1}{qx} (q_\alpha x_\nu + q_\nu x_\alpha)\right) \right. \right. q_\beta
\nonumber \\ && -
         \left(\varepsilon_\mu - q_\mu \frac{\varepsilon x}{q x}\right)\left(g_{\beta \nu} -
        \frac{1}{qx} (q_\beta x_\nu + q_\nu x_\beta)\right) q_\alpha
\nonumber \\ && -
         \left(\varepsilon_\nu - q_\nu \frac{\varepsilon x}{q x}\right)\left(g_{\alpha \mu} -
        \frac{1}{qx} (q_\alpha x_\mu + q_\mu x_\alpha)\right) q_\beta
\nonumber \\ &&+
         \left. \left(\varepsilon_\nu - q_\nu \frac{\varepsilon x}{q.x}\right)\left( g_{\beta \mu} -
        \frac{1}{qx} (q_\beta x_\mu + q_\mu x_\beta)\right) q_\alpha \right]
   \int {\cal D}\alpha_i e^{i (\alpha_{\bar q} + v \alpha_g) qx} {\cal T}_1(\alpha_i)
\nonumber \\ &&+
        \left[\left(\varepsilon_\alpha - q_\alpha \frac{\varepsilon x}{qx}\right)
        \left(g_{\mu \beta} - \frac{1}{qx}(q_\mu x_\beta + q_\beta x_\mu)\right) \right. q_\nu
\nonumber \\ &&-
         \left(\varepsilon_\alpha - q_\alpha \frac{\varepsilon x}{qx}\right)
        \left(g_{\nu \beta} - \frac{1}{qx}(q_\nu x_\beta + q_\beta x_\nu)\right)  q_\mu
\nonumber \\ && -
         \left(\varepsilon_\beta - q_\beta \frac{\varepsilon x}{qx}\right)
        \left(g_{\mu \alpha} - \frac{1}{qx}(q_\mu x_\alpha + q_\alpha x_\mu)\right) q_\nu
\nonumber \\ &&+
         \left. \left(\varepsilon_\beta - q_\beta \frac{\varepsilon x}{qx}\right)
        \left(g_{\nu \alpha} - \frac{1}{qx}(q_\nu x_\alpha + q_\alpha x_\nu) \right) q_\mu
        \right]
    \int {\cal D} \alpha_i e^{i (\alpha_{\bar q} + v \alpha_g) qx} {\cal T}_2(\alpha_i)
\nonumber \\ &&+
        \frac{1}{qx} (q_\mu x_\nu - q_\nu x_\mu)
        (\varepsilon_\alpha q_\beta - \varepsilon_\beta q_\alpha)
    \int {\cal D} \alpha_i e^{i (\alpha_{\bar q} + v \alpha_g) qx} {\cal T}_3(\alpha_i)
\nonumber \\ &&+
        \left. \frac{1}{qx} (q_\alpha x_\beta - q_\beta x_\alpha)
        (\varepsilon_\mu q_\nu - \varepsilon_\nu q_\mu)
    \int {\cal D} \alpha_i e^{i (\alpha_{\bar q} + v \alpha_g) qx} {\cal T}_4(\alpha_i)
                        \right\},\nonumber~~~~~~~~~(A.1)
\end{eqnarray}
where, $\chi$ is the magnetic susceptibility of the quarks,
$\varphi_\gamma(u)$ is the leading twist 2, $\psi^v(u)$,
$\psi^a(u)$, ${\cal A}$ and ${\cal V}$ are the twist 3 and
$h_\gamma(u)$, $\mathbb{A}$, ${\cal T}_i$ ($i=1,~2,~3,~4$) are the
twist 4 photon DA's, respectively.  The measure ${\cal D} \alpha_i$ is defined as
\begin{eqnarray}
\int {\cal D} \alpha_i = \int_0^1 d \alpha_{\bar q} \int_0^1 d
\alpha_q \int_0^1 d \alpha_g \delta(1-\alpha_{\bar
q}-\alpha_q-\alpha_g).\nonumber\hskip 2.9cm (A.2)
\end{eqnarray}

The explicit expressions of the photon distribution amplitudes (DA's) with different twists are \cite{Ball}:
\begin{eqnarray}
\varphi_\gamma(u) &=& 6 u \bar u \left( 1 + \varphi_2(\mu)
C_2^{\frac{3}{2}}(u - \bar u) \right),
\nonumber \\
\psi^v(u) &=& 3 \left(3 (2 u - 1)^2 -1 \right)+\frac{3}{64} \left(15
w^V_\gamma - 5 w^A_\gamma\right)
                        \left(3 - 30 (2 u - 1)^2 + 35 (2 u -1)^4
                        \right),
\nonumber \\
\psi^a(u) &=& \left(1- (2 u -1)^2\right)\left(5 (2 u -1)^2 -1\right)
\frac{5}{2}
    \left(1 + \frac{9}{16} w^V_\gamma - \frac{3}{16} w^A_\gamma
    \right),
\nonumber \\
{\cal A}(\alpha_i) &=& 360 \alpha_q \alpha_{\bar q} \alpha_g^2
        \left(1 + w^A_\gamma \frac{1}{2} (7 \alpha_g - 3)\right),
\nonumber \\
{\cal V}(\alpha_i) &=& 540 w^V_\gamma (\alpha_q - \alpha_{\bar q})
\alpha_q \alpha_{\bar q}
                \alpha_g^2,
\nonumber \\
h_\gamma(u) &=& - 10 \left(1 + 2 \kappa^+\right) C_2^{\frac{1}{2}}(u
- \bar u),
\nonumber \\
\mathbb{A}(u) &=& 40 u^2 \bar u^2 \left(3 \kappa - \kappa^+
+1\right) \nonumber \\ && +
        8 (\zeta_2^+ - 3 \zeta_2) \left[u \bar u (2 + 13 u \bar u) \right.
\nonumber \\ && + \left.
                2 u^3 (10 -15 u + 6 u^2) \ln(u) + 2 \bar u^3 (10 - 15 \bar u + 6 \bar u^2)
        \ln(\bar u) \right],
\nonumber \\
{\cal T}_1(\alpha_i) &=& -120 (3 \zeta_2 + \zeta_2^+)(\alpha_{\bar
q} - \alpha_q)
        \alpha_{\bar q} \alpha_q \alpha_g,
\nonumber \\
{\cal T}_2(\alpha_i) &=& 30 \alpha_g^2 (\alpha_{\bar q} - \alpha_q)
    \left((\kappa - \kappa^+) + (\zeta_1 - \zeta_1^+)(1 - 2\alpha_g) +
    \zeta_2 (3 - 4 \alpha_g)\right),
\nonumber \\
{\cal T}_3(\alpha_i) &=& - 120 (3 \zeta_2 - \zeta_2^+)(\alpha_{\bar
q} -\alpha_q)
        \alpha_{\bar q} \alpha_q \alpha_g,
\nonumber \\
{\cal T}_4(\alpha_i) &=& 30 \alpha_g^2 (\alpha_{\bar q} - \alpha_q)
    \left((\kappa + \kappa^+) + (\zeta_1 + \zeta_1^+)(1 - 2\alpha_g) +
    \zeta_2 (3 - 4 \alpha_g)\right),\nonumber \\
{\cal S}(\alpha_i) &=& 30\alpha_g^2\{(\kappa +
\kappa^+)(1-\alpha_g)+(\zeta_1 + \zeta_1^+)(1 - \alpha_g)(1 -
2\alpha_g)\nonumber \\&+&\zeta_2
[3 (\alpha_{\bar q} - \alpha_q)^2-\alpha_g(1 - \alpha_g)]\},\nonumber \\
\tilde {\cal S}(\alpha_i) &=&-30\alpha_g^2\{(\kappa -
\kappa^+)(1-\alpha_g)+(\zeta_1 - \zeta_1^+)(1 - \alpha_g)(1 -
2\alpha_g)\nonumber \\&+&\zeta_2 [3 (\alpha_{\bar q} -
\alpha_q)^2-\alpha_g(1 - \alpha_g)]\}.\nonumber~~~~~~~~~~~~~~~~~~~~~~~~~~~~~~~~~~~~~~~(A.3)
\end{eqnarray}
The constants appearing in the above wave functions are given as
\cite{Ball}: $\varphi_2(1~GeV) = 0$, $w^V_\gamma = 3.8 \pm 1.8$,
$w^A_\gamma = -2.1 \pm 1.0$, $\kappa = 0.2$, $\kappa^+ = 0$,
$\zeta_1 = 0.4$, $\zeta_2 = 0.3$, $\zeta_1^+ = 0$ and $\zeta_2^+ =
0$.

\section*{Appendix B}
In this appendix, we present the  explicit expressions for the
functions, $F_{1}(0)$, $F_{2}(0)$, $F_{3}(0)$ and $F_{4}(0)$.
\begin{eqnarray}\label{F1}
 F_{1}(q^2=0)&=&-\frac{1}{2m_{\Delta}\lambda_{\Delta}^{2}}e^{\frac{m_{\Delta}^{2}}{M^{2}}}\left(\vphantom{\int_0^{x_2}}\right.e_{u}\left\{\vphantom{\int_0^{x_2}}\right.\langle \bar d d\rangle\left[\vphantom{\int_0^{x_2}}\right.\frac{M^2[12E_{1}(x)M^2-5E_{0}(x)m_{0}^2]}{27\pi^2}
\nonumber\\&-&\frac{2}{81}f_{3\gamma}\{54E_{0}(x)M^{2}-15m_{0}^2\}\psi^{v}(u_{0})\left.\vphantom{\int_0^{x_2}}\right]
\nonumber\\&+&
\langle \bar uu\rangle\left[\vphantom{\int_0^{x_2}}\right.\frac{M^2[3E_{1}(x)M^2(8-15\zeta_{1})-10E_{0}(x)m_{0}^2]}{27\pi^2}
\nonumber\\&-&\frac{2}{81}f_{3\gamma}\{54E_{0}(x)M^{2}-15m_{0}^2\}\psi^{v}(u_{0})\left.\vphantom{\int_0^{x_2}}\right]\left.\vphantom{\int_0^{x_2}}\right\}+e_{d}\left\{\vphantom{\int_0^{x_2}}\right.\langle \bar d d\rangle\left[\vphantom{\int_0^{x_2}}\right.-\frac{5E_{1}(x)M^4\zeta_{1}}{12\pi^2}\left.\vphantom{\int_0^{x_2}}\right]\nonumber\\&+&\langle \bar uu\rangle\left[\vphantom{\int_0^{x_2}}\right.\frac{M^2[12E_{1}(x)M^2-5E_{0}(x)m_{0}^2]}{27\pi^2}
\nonumber\\&-&\frac{2}{81}f_{3\gamma}\{54E_{0}(x)M^{2}-15m_{0}^2\}\psi^{v}(u_{0})\left.\vphantom{\int_0^{x_2}}\right]\left.\vphantom{\int_0^{x_2}}\right\}\left.\vphantom{\int_0^{x_2}}\right),\nonumber~~~~~~~~~~~~~~~~~~~~~~~~~~~(B.1)
\end{eqnarray}
\begin{eqnarray}\label{F2}
 F_{2}(q^2=0)&=&-\frac{m_{\Delta}}{\lambda_{\Delta}^{2}}e^{\frac{m_{\Delta}^{2}}{M^{2}}}
\left(\vphantom{\int_0^{x_2}}\right.e_{u}\left\{\vphantom{\int_0^{x_2}}\right.\langle
\bar d
d\rangle\left[\vphantom{\int_0^{x_2}}\right.\frac{5m_{0}^2-10E_{0}(x)M^2}{216\pi^2}
\nonumber\\&+&\frac{f_{3\gamma}}{324M^{2}}\{11m_{0}^2-36M^{2}\}\psi^{a}(u_{0})\left.\vphantom{\int_0^{x_2}}\right]\nonumber\\&+&\langle
\bar
uu\rangle\left[\vphantom{\int_0^{x_2}}\right.\frac{1}{216\pi^2}[5m_{0}^2+6E_{0}(x)M^2(-3+2\eta_{3}
+6\eta_{4}-8\eta_{5}+2\eta_{6}-4\eta_{7}-8\eta_{8}
\nonumber\\&-&2\eta_{9}+4\eta_{10}+24\zeta_{3})-9E_{0}(x)M^2\mathbb{A}(u_{0}+12E_{1}(x)M^4\chi\varphi_{\gamma}(u_{0})]
\nonumber\\&+&
\frac{f_{3\gamma}}{324M^{2}}\{11m_{0}^2-36M^{2}\}\psi^{a}(u_{0})\left.\vphantom{\int_0^{x_2}}\right]\left.\vphantom{\int_0^{x_2}}\right\}\nonumber\\&+&
e_{d}\left\{\vphantom{\int_0^{x_2}}\right.\langle \bar
uu\rangle\left[\vphantom{\int_0^{x_2}}\right.\frac{5m_{0}^2-18E_{0}(x)M^2}{216\pi^2}
\nonumber\\&+&\frac{f_{3\gamma}}{324M^{2}}\{11m_{0}^2-36M^{2}\}\psi^{a}(u_{0})\left.\vphantom{\int_0^{x_2}}\right]
\nonumber\\&+&\langle \bar
dd\rangle\left[\vphantom{\int_0^{x_2}}\right.\frac{M^2}{144\pi^2}[4E_{0}(x)(\eta_{3}
+3\eta_{4}-4\eta_{5}+\eta_{6}-2\eta_{7}-4\eta_{8}
\nonumber\\&-&\eta_{9}+2\eta_{10}+12\zeta_{3})-3E_{0}(x)M^2\mathbb{A}(u_{0}+4E_{1}(x)M^4\chi\varphi_{\gamma}(u_{0})]
+\left.\vphantom{\int_0^{x_2}}\right]\left.\vphantom{\int_0^{x_2}}\right\}\left.\vphantom{\int_0^{x_2}}\right),\nonumber(B.2)
\end{eqnarray}
\begin{eqnarray}\label{F3}
 F_{3}(q^2=0)&=&-\frac{2m_{\Delta^{2}}}{\lambda_{\Delta}^{2}}e^{\frac{m_{\Delta}^{2}}{M^{2}}}
\left(\vphantom{\int_0^{x_2}}\right.e_{u}\left\{\vphantom{\int_0^{x_2}}\right.\frac{-E_{1}(x)M^4(u_{0}-1)u_{0}}{12\pi^4}
\nonumber\\&+&\frac{f_{3\gamma}E_{0}(x)M^2}{18\pi^2}\left[\vphantom{\int_0^{x_2}}\right.10\eta_{1}
-8\eta_{11}-8\eta_{2}+4\zeta_{2}(1-2u_{0})-\psi^{a}(u_{0})\left.\vphantom{\int_0^{x_2}}\right]\nonumber\\&-&
\frac{4}{81M^{4}}\langle \bar uu\rangle(\langle \bar
uu\rangle+\langle \bar
dd\rangle)\left[\vphantom{\int_0^{x_2}}\right.
-10m_{0}^2\zeta_{3}+6M^{2}(6\zeta_{3}+3\xi_{1}+4\xi_{2}-\xi_{3})\left.\vphantom{\int_0^{x_2}}\right]\left.\vphantom{\int_0^{x_2}}\right\}\nonumber\\&+&e_{d}\left\{\vphantom{\int_0^{x_2}}\right.\frac{-E_{1}(x)M^4(u_{0}-1)u_{0}}{24\pi^4}
\nonumber\\&+&\frac{f_{3\gamma}E_{0}(x)M^2}{36\pi^2}\left[\vphantom{\int_0^{x_2}}\right.10\eta_{1}
-8\eta_{11}-8\eta_{2}+4\zeta_{2}(1-2u_{0})-\psi^{a}(u_{0})\left.\vphantom{\int_0^{x_2}}\right]\nonumber\\&-&
\frac{4}{81M^{4}}\langle \bar uu\rangle\langle \bar
dd\rangle\left[\vphantom{\int_0^{x_2}}\right.
-10m_{0}^2\zeta_{3}+6M^{2}(6\zeta_{3}+3\xi_{1}+4\xi_{2}-\xi_{6})\left.\vphantom{\int_0^{x_2}}\right]\left.\vphantom{\int_0^{x_2}}\right\}\left.\vphantom{\int_0^{x_2}}\right),\nonumber~~~~(B.3)
\end{eqnarray}
\begin{eqnarray}\label{F4}
 F_{4}(q^2=0)&=&-\frac{4m_{\Delta^{2}}}{\lambda_{\Delta}^{2}}e^{\frac{m_{\Delta}^{2}}{M^{2}}}
\left(\vphantom{\int_0^{x_2}}\right.e_{u}\left\{\vphantom{\int_0^{x_2}}\right.\frac{-E_{1}(x)M^4(u_{0}-1)u_{0}^2}{8\pi^4}
\nonumber\\&+&\frac{f_{3\gamma}E_{0}(x)M^2}{18\pi^2}\left[\vphantom{\int_0^{x_2}}\right.
8(\eta_{11}+\eta_{2})-12\zeta_{2}u_{0}^2+8u_{0}(\eta_{1}-3\eta_{11}-\eta_{2}+\zeta_{2})
\nonumber\\&+&(u_{0}-2)u_{0}\psi^{a}(u_{0})\left.\vphantom{\int_0^{x_2}}\right]\nonumber\\&-&
\frac{4}{81M^{4}}\langle \bar uu\rangle(\langle \bar
uu\rangle+\langle \bar
dd\rangle)\left[\vphantom{\int_0^{x_2}}\right.
-10m_{0}^2u_{0}\zeta_{3}+12M^{2}(1+3u_{0}\zeta_{3}+\xi_{2}-\xi_{3})\left.\vphantom{\int_0^{x_2}}\right]\left.\vphantom{\int_0^{x_2}}\right\}\nonumber\\&+&e_{d}\left\{\vphantom{\int_0^{x_2}}\right.\frac{-E_{1}(x)M^4(u_{0}-1)u_{0}^2}{16\pi^4}
\nonumber\\&+&\frac{f_{3\gamma}E_{0}(x)M^2}{36\pi^2}\left[\vphantom{\int_0^{x_2}}\right.
8(\eta_{11}+\eta_{2})-12u_{0}^2\zeta_{2}+8u_{0}(\eta_{1}-3\eta_{11}-\eta_{2}+\zeta_{2})\nonumber\\&+&(u_{0}-2)u_{0}\psi^{a}(u_{0})\left.\vphantom{\int_0^{x_2}}\right]\nonumber\\&+&
\frac{4}{81M^{4}}\langle \bar uu\rangle\langle \bar
dd\rangle\left[\vphantom{\int_0^{x_2}}\right.
10m_{0}^2u_{0}\zeta_{3}-12M^{2}(3u_{0}\zeta_{3}+\xi_{2}-\xi_{3})\left.\vphantom{\int_0^{x_2}}\right]\left.\vphantom{\int_0^{x_2}}\right\}\left.\vphantom{\int_0^{x_2}}\right),\nonumber~~~~~~~~(B.4)
\end{eqnarray}
where, the functions entering the above equations are given as
\begin{eqnarray}\label{etalar}
\eta_{i} &=& \int {\cal D}\alpha_i \int_0^1 dv f_{i}(\alpha_i)
\delta(\alpha_{ q} + v \alpha_g -  u_0),
\nonumber \\
\xi_{i} &=& \int {\cal D}\alpha_i \int_0^1 d\bar v g_{i}(\alpha_i)
\theta(\alpha_{ q} + v \alpha_g -  u_0),
\nonumber \\
\zeta_{i} &=&  \int_{u_{0}}^1 du h_{i}(u),
\nonumber \\
E_{n}(x)&=&1-e^{-x}\sum_{k=0}^{n}\frac{x^{k}}{k!},\nonumber ~~~~~~~~~~~~~~~~~~~~~~~~~~~~~~~~~~~~~~~~~~~~~~~~~~~~~~~(B.5)
\end{eqnarray}
 and  $f_{1}(\alpha_i)={\cal A}(\alpha_i)$, $f_{2}(\alpha_i)=v{\cal A}(\alpha_i)$, $f_{3}(\alpha_i)={\cal S}(\alpha_i)$, $f_{4}(\alpha_i)=\tilde{\cal S}(\alpha_i)$, $f_{5}(\alpha_i)=v\tilde{\cal S}(\alpha_i)$, $f_{6}(\alpha_i)=g_{2}(\alpha_i)={\cal T}_{2}(\alpha_i)$, $f_{7}(\alpha_i)=v{\cal T}_{2}(\alpha_i)$, $f_{8}(\alpha_i)=v{\cal T}_{3}(\alpha_i)$, $f_{9}(\alpha_i)=g_{3}(\alpha_i)={\cal T}_{4}(\alpha_i)$, $f_{10}(\alpha_i)=v{\cal T}_{4}(\alpha_i)$, $f_{11}(\alpha_i)=v{\cal V}(\alpha_i)$, $g_{1}(\alpha_i)={\cal T}_{1}(\alpha_i)$, $h_{1}(u)=h_{\gamma}(u)$, $h_{2}(u)=\psi^{v}(u)$ and $h_{3}(u)=(u-u_{0})h_{\gamma}(u)$ are the photon distribution amplitudes. Note that, in the above equations, $x=s_{0}/M^{2}$, $\bar v=1-v$ and the Borel parameter $M^2$  is defined as $M^{2}=\frac{M_{1}^{2}M_{2}^{2}}{M_{1}^{2}+M_{2}^{2}}$ and
$u_{0}=\frac{M_{1}^{2}}{M_{1}^{2}+M_{2}^{2}}$.  Since the masses of
the initial and final baryons are the same, we will set
$M_{1}^{2}=M_{2}^{2}$ and $u_{0}=1/2$.
\end{document}